\documentclass[aps,showpacs,prl,twocolumn,groupedaddress]{revtex4}
\usepackage{mathptmx}
\usepackage{graphicx}
\usepackage{dcolumn}
\usepackage{bm}
\usepackage{color}

\begin{document}

\title{Origin of Unexpected Low Energy Structure in Photoelectron Spectra \\ Induced by Mid-Infrared Strong Laser Fields}
\author{Chengpu Liu}
\email{Chengpu.Liu@mpi-hd.mpg.de}
\affiliation{Max-Planck-Institut f\"ur Kernphysik, Saupfercheckweg 1, D-69117 Heidelberg, Germany}

\author{Karen Z. Hatsagortsyan}
\email{k.hatsagortsyan@mpi-hd.mpg.de}
\affiliation{Max-Planck-Institut f\"ur Kernphysik, Saupfercheckweg 1, D-69117 Heidelberg, Germany}

\date{\today}

\begin{abstract}
Using a semiclassical model which
incorporates tunneling and Coulomb field effects, the origin of the unexpected low-energy structure (LES) in above-threshold ionization spectrum
observed in recent experiments [C. I. Blaga \textit{et al.}, Nature Phys. {\bf 5}, 335 (2009) and W. Quan \emph{et al.}, Phys. Rev. Lett. {\bf 103}, 093001 (2009)] is
identified. We show that the LES arises due to an interplay between multiple
forward scattering of an ionized electron and the
electron momentum disturbance by the Coulomb field immediately after the ionization.
The multiple forward scattering is mainly responsible for the appearance of LES, while the initial disturbance
mainly determines the position of the LES peaks.
The scaling laws for the LES parameters, such as the contrast ratio and the maximal energy, versus the laser
intensity and wavelength are deduced.
\end{abstract}

\maketitle

The physics of strong-field photoionization has been extensively
studied in the last two decades and is assumed to be
well-understood \cite{Becker_review} with the three-step model
\cite{CorkumPRL93} playing a fundamental role. The attosecond science has been developed based on the knowledge how to control the motion of electrons on the atomic scale 
which paved a way for new methods for measuring molecular structure \cite{Lein}.
The strong field approximation (SFA) \cite{SFA} serves as an analytical tool to understand the strong-field ionization phenomena which
can describe the rescattering via a perturbative approach \cite{Becker_review}.
Recently, however, a number of strong-field ionization phenomena has been revealed which are caused by non-perturbative influence of the Coulomb field of the atomic core \cite{Bauer}:
double-hump and interference structures \cite{UllrichPRL03,interference_structures} in the
momentum distribution of photoelectrons near ionization threshold, frustrated tunneling ionization
\cite{EichmannNature09}, and multiphoton assisted recombination \cite{Gallagher}.
With the advent and further 
improvement of  intense femtosecond mid-infrared laser sources \cite{MidInfrared09},
the classical regime of strong-field ionization, when the Keldysh parameter $\gamma\ll 1$ and one expects the SFA to provide an adequate description,
has been subjected to more attentive investigation. Here, $\gamma = \sqrt{I_p/2U_p}$, $I_p$ is the ionization potential, and \emph{U}$_p$ the ponderomotive energy.
The recent two experiments by C. Blaga et al. \cite{DiMauroNP09} and
W. Quan et al. \cite{XuPRL09}
on the photoionization of atoms and molecules in strong
mid-infrared laser fields reveal a previously
unexpected characteristic spike-like low-energy structure (LES) in
the energy distribution of electrons emitted along the laser
polarization direction, see Fig. 1 (a).
These observations manifest a striking contrast to the prediction of the SFA
and point to a lack of complete understanding of strong field
physics. The numerical solutions of the 
Schr\"odinger equation  within the single active electron
approximation can successfully reproduces the measured LES in the
case of any atomic potential. However, the 
calculations using 
SFA or SFA with Coulomb corrections \cite{KFR2} fail to describe the LES.
Varying the laser polarization from linear to circular, LES
is significantly reduced. The latter indicates that
forward rescattering \cite{PaulusPRL94} is
playing an essential role in this process which has been pointed
out by Blaga et al. \cite{DiMauroNP09} and Faisal \cite{FaisalNP09}.
However, the mechanism which creates the LES remains obscure. Many questions remain unanswered: How  exactly does the LES arise?
Why does it have a peaked structure? Why is the effect of rescattering more pronounced in mid-infrared laser fields?

In this Letter, we investigate in detail and identify the mechanism of LES.
We employ the classical-trajectory Monte Carlo (CTMC) method with tunneling and the Coulomb field of the atomic core fully taken into account. In addition, we provide a qualitative theoretical estimation
for the Coulomb field effects: initial Coulomb focusing (CF),
multiple forward scattering and asymptotic CF.
We quantify their relative role in the electron dynamics and conclude that
1) the behavior of the transverse (with respect to the laser polarization direction) momentum change of the electron due
to  Coulomb field effects with respect to the ionization phase is the key for understanding of the LES;
2) at mid-infrared wavelengths, multiple scattering of the
ionized electron plays a decisive non-perturbative role.
In particular, the transverse momentum change of the electron due
to  multiple scattering distorts the electron phase space to create peaks at low electron energies, while the longitudinal momentum change due to  initial CF shifts the peak energy to higher energies.
We investigate and explain qualitatively the scaling of the LES parameters.

In the $\gamma \ll 1$ regime, the electron oscillation amplitude
in the laser field $\alpha=E_0/\omega^2$ exceeds the distance of the
electron from the atomic core at the tunnel exit $z_0=I_p/E_0$:
$\alpha/z_0\sim 2/\gamma^2 \gg 1$ and the transversal distance
traveled by the electron during one laser period $x_0\sim 2\pi \sqrt{E_0}/(2I_p)^{1/4}\omega$:
$\alpha/x_0\sim (1/2\pi)\sqrt{2I_p/\gamma\omega}\gg 1$, where $E_0$ and
$\omega$ are the laser field amplitude and frequency, respectively (atomic
units are used throughout). Therefore, in this regime, the
electron travels far from the core during its oscillation in the
laser field, and the Coulomb field distorts the electron trajectory
only at positions very close to the core. This happens immediately
after ionization, corresponding to the initial CF, and when the electron revisits the atomic
core and re-scatters.
The number of scatterings is large for mid-infrared wavelengths and low-energy photoelectrons \cite{N_s}: $N_s\sim \alpha/x_0\sim
10$ at the parameters in \cite{DiMauroNP09}. The third Coulomb
effect is the asymptotic CF when the electron momentum is disturbed by
the Coulomb field after the laser pulse is switched off. This usually
plays an important role for low-energy photoelectrons
\cite{EichmannNature09,BurgPRA04,RussiaLP09} but is not essential for the LES as shown below.

\begin{figure}[t]
\includegraphics[width=9 cm]{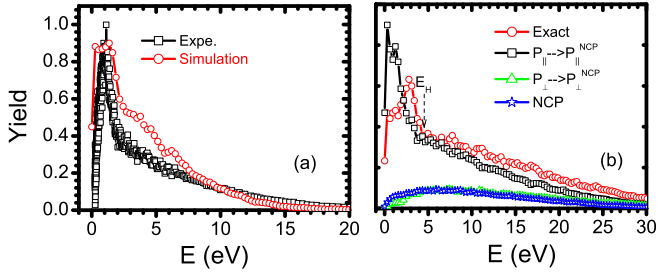}
\caption{(color online). Photoelectron spectra (PES): (a)
The experimental result (squares) for a Xenon atom in a laser field with
peak intensity $I_0 = 3.2 \times 10^{13} W/cm^{2}$ and wavelength
 $\lambda = 2.3\mu m$ \cite{DiMauroNP09} as well as the corresponding
CTMC-simulation (circles). (b) CTMC-simulations for a hydrogen atom
with $\lambda = 2\mu m$, $I_0 = 9.0 \times 10 ^{13}W/cm ^2$: exact
(circles);  with totally neglecting the Coulomb potential (NCP)
(stars); NCP only for the electron longitudinal momenta
$P_{\parallel}$ (squares) and  NCP  only for the transverse momenta
$P_{\bot}$ (triangles). The high-energy limit of LES defined by the
break in slope is indicated with an arrow.} \label{Fig_1}
\end{figure}

In our 3D CTMC simulation, the ionized electron wave packet is
formed according to the ADK ionization
rate \cite{ADK} and further propagates classically. 
The electrons are born at the tunnel exit with the following conditions: (i) Along the laser
polarization direction, the initial position $z_{i}$ is derived from
the effective potential theory \cite{Landau77} and the initial momentum
$p_{i\parallel} = 0$ \cite{Brabec}; (ii) The transversal coordinates are $x_{i} = y_{i} = 0
$. The transverse momentum $p_{i\bot}$ follows the
corresponding ADK distribution \cite{ADK}. The transverse momentum components are $p_{ix} = p_{i\bot}\sin\phi$ and $p_{iy} =
p_{i\bot}\cos\phi$, where $\phi$ is the azimuthal angle randomly distributed within
an interval of $(0, 2\pi)$. The positions and momenta of
electrons after the laser pulse are used to calculate the final
asymptotic momenta \cite{RussiaLP09} at the detector. Only
electrons emitted along the laser polarization direction
within an angle of $\pm 2.5^{\circ}$ are collected. The
laser pulse is half-trapezoidal, constant for the first ten cycles
and ramped off within the last three cycles. The electrons are
launched within the first half cycle ($\omega t_{i}\in [0,\pi]$). Our model
provides an adequate description of the photoelectron spectrum (PES)
and is qualitatively consistent with the experimental results as
an example in Fig. \ref{Fig_1}(a) shows.
\begin{figure}[t]
\includegraphics[width=8.5 cm]{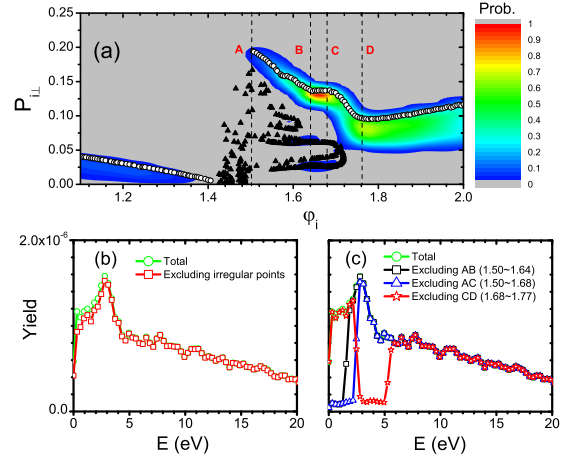}
\caption{(color online).
(a) The distribution of electrons in the LES in phase space (the initial transverse momentum versus the ionization phase $\varphi_i \equiv\omega t_{i}$) within the electron energy interval $(0,20)$ eV, with color coded probability.
Irregular points are shown by triangles. The circles indicate the maximum
probability for each phase. The laser field is maximal at $\varphi_i=\pi/2$.
(b) PES with and without irregular
points. (c) PES, the electrons born in the specified phase range
are removed. The laser and atom parameters are the same as
in Fig. 1(b).} \label{Fig_2}
\end{figure}

Now we turn to the clarification of the physical mechanism behind
the LES. For simplicity, we consider hydrogen atoms because the atomic
structure is not essential for LES.
Firstly, we investigate the changes in PES when some factors are neglected,
see Fig. \ref{Fig_1} (b). If the Coulomb potential is neglected
after ionization, the LES disappears, as expected. If the
Coulomb field effect  is neglected only on the longitudinal momentum, the LES
shifts towards lower energy and enhances. However, if
the Coulomb field effect  is neglected only on the transverse momentum, the LES
disappears completely. We can deduce that the change of the transverse momentum
due to the Coulomb potential is the main source of LES at which we
look more closely next.

Secondly, we investigate the phase space distribution of electrons
which contribute to the LES, the initial transverse momentum
versus the birth phase shown in Fig. \ref{Fig_2}(a): (i) The
electron ionized near the peak of the laser field with a low
transverse momentum has a low drift momentum, scatters many times
by the atomic core with a low impact parameter and, as a result,
shows chaotic behavior (the triangles in Fig. \ref{Fig_2}(a)).
However, the electrons with chaotic dynamics mostly contribute to
the high-energy part of PES essential for the formation
of a plateau in above-threshold ionization. 
The contribution of irregular points to the LES amounts only to a few percent
\cite{chaotic}. For this reason, when these points are discarded, the LES in PES shows
only a little reduction in its lower energy part, see Fig.
\ref{Fig_2}(b). Therefore, the chaotic dynamics is not the cause
of the LES.
(ii) A large number of electrons
with a large initial transverse momentum ($p_{i\bot}$)
is concentrated after the interaction in the phase space with a
low transverse momentum. This is usually termed as CF \cite{IvanovPRA01}. The CF
is largest near the laser peak and decreases with
the increasing of the ionization phase. (iii) More
importantly,   the decrease of $p_{i\bot}$ is not monotonic but
shows a step-like slope change. We state that this slope
change is responsible for the appearance of peaks in the LES. In fact, near the points of the slope change of the $p_{i\bot}(\varphi_i)$, the phase space of electrons  contributing to the LES (per unit $\varphi_i$) has local maxima. This
can be tested via artificially discarding the electrons between
certain characteristic phases, see Fig. \ref{Fig_2}(c).
The latter shows that the occurrence of the highest part of LES
is, in fact, related to the electrons with ionization phases within
an interval $(1.64,1.77)$  exactly corresponding to the
slope change between B and D.
The behavior of $p_{i\bot}$ versus the ionization phase, in fact, reflects the one of the
transverse momentum change due to the Coulomb field ($\delta
p_{\bot}$), see Fig. \ref{Fig_4}(a) below. Then, one needs to explain $\delta
p_{\bot}$  non-monotonic behavior with respect to
$\phi_i$.
\begin{figure}[b]
\includegraphics[width=9 cm]{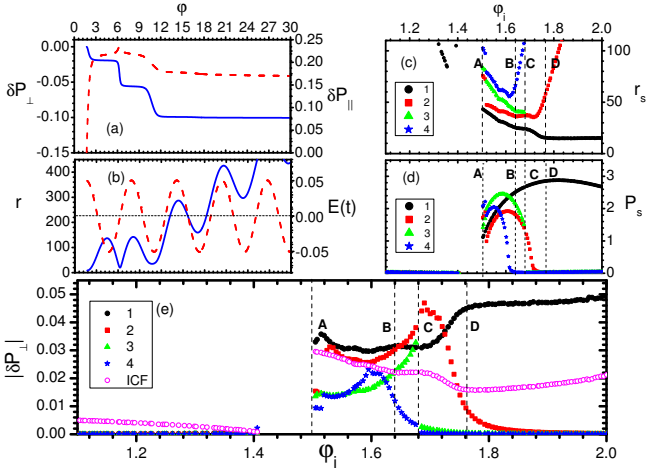}
\caption{(color online). (a) The transverse $\delta P_{\bot}$
(solid) and longitudinal $\delta P_{\parallel}$ (dashed) momentum
changes due to the Coulomb field. (b) The distance $r$ (solid) from the core and
electric field (dashed) vs. the laser phase for
one specific trajectory ($\varphi_i = 1.72$). (c) The position
$r_s$ and (d) the momentum $p_s$  at the $s$th scattering versus the
ionization phase. (e) The estimated $\delta P_{\bot}$
due to  different number of 
scatterings and initial CF. The
laser and atom parameters are the same as in Fig. 1(b).}
\label{Fig_3}
\end{figure}
\begin{figure}[t]
\includegraphics[width=9 cm]{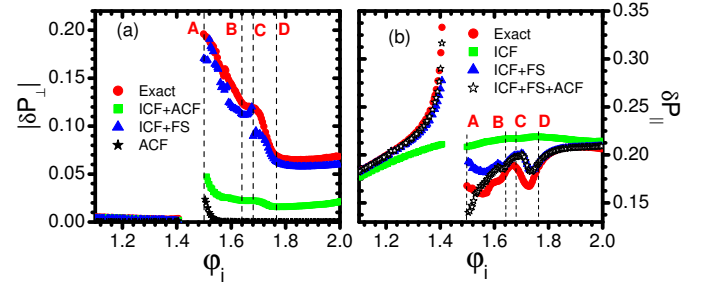}
\caption{(color online). The estimation of (a) $|\delta P
_{\bot}|$  and (b) $\delta P _{\parallel}$ described in the text. The contributions of asymptotic CF, the
combined contribution of asymptotic and initial CF, and the one of asymptotic CF and scattering are
shown. The laser and atom parameters are the same as
in Fig. 1(b).} \label{Fig_4}
\end{figure}

Thirdly, we investigate the transverse and longitudinal
momentum disturbance of the electron  due to the Coulomb
field ($\delta p_{\bot}$ and
$\delta p_{\parallel}$). 
For one specific trajectory
contributing to LES, $\delta p_{\bot}$ and
$\delta p_{\parallel}$ are shown in Fig. \ref{Fig_3}(a). After the electron's birth, $\delta
p_{\bot,\parallel}$ changes abruptly during a fractional part of
the laser period which is known as initial CF
\cite{IvanovPRA01}. Further momentum change takes
place due to scattering when the electron passes the core at the minimum
distance $r$ (Fig. \ref{Fig_3}(b)). We proceed to estimate the
contributions of  initial CF and multiple scattering, respectively, for electrons
with the maximal probability for each phase (circles in Fig.
2(a)). (i) The transverse momentum change due to Coulomb
potential $V(r)$ at the $s{\rm th}$ scattering can be estimated as
$\delta p_{\bot}^{(s)} \approx \int \nabla_{\bot}V(r(t))dt\sim
(\rho_s/r_s^3) \delta t_s$, where $r_s$ is the distance from the
core at the scattering moment, $\rho_s$ is the one in the transversal plane,
and $\delta t_s$ is the scattering duration. When the electron velocity
$p_s$ at scattering is large, $\delta t_s\sim 2r_s/p_s$.
In the opposite case, $\delta t_s\sim 2
\sqrt{2r_s/|E(\varphi_s)|}$ is determined by the laser field
$E(\varphi_s)$ at scattering. Accordingly,
$\delta p_{\bot}^{(s)} = -{2\rho_s}/{(r_s^2 p_s)}$, 
if $p_s^{2} \gg r_s |E(\varphi_s)|$, or 
$\delta p_{\bot}^{(s)} = -{2 ^{3/2} \rho_s}/{
\sqrt{r_s^5|E(\varphi_s)|}}$, otherwise.
$\delta p_{\bot}^{(s)}$ is sensitive to $r_s$ and $p_s$. Their  values for
the different scattering  events are shown in Figs. \ref{Fig_3}(c) and (d). As
for the formula of the $\delta p^{(s)}_{\parallel}$, $\rho_s$ in
the expression for $\delta p_{\bot}^{(s)}$ should be substituted by the absolute value
of the scattering  coordinate $|z_s|$ along the laser polarization direction.
(ii) The transverse momentum change due to initial CF can be estimated as $\delta p_{\bot}^{(I)}
=-2p_{i\bot}f_{\bot}(E_0,\omega)|E(\varphi_i)|/(2I_p)^2$ and the longitudinal one as
$\delta p_{\parallel}^{(I)} = \pi f_{\parallel}(E_0,\omega)|E(\varphi_i)|/(2I_p)^{3/2}$, where the functions $f_{\bot,\parallel}(E_0,\omega)$
describe the deviation from the simple estimate of \cite{RussiaLP09}.
(iii) We estimate the asymptotic CF contribution comparing the asymptotic
electron momentum with the one after switching off the laser pulse.

In Fig. 3 (e), the relative contribution of multiple scattering   (up to 4th scattering) and initial CF to $\delta p_{\bot}$ is shown calculated using the
above formulas. At $\varphi>1.77$ (line D), mainly 1st scattering and initial CF
plays a role. With a little shift of phase towards 1.68 (line C),
the 2nd scattering enhances rapidly due to the increase of the scattering 
energy and the 1st scattering weakens. The significant
contribution from the 2nd scattering results in the slope change between
lines C and D in Fig. 2 (b).
Decreasing the phase further, the 3rd scattering becomes comparable with
initial CF, the 1st and 2nd scattering. The competition between 1st-3rd scattering and initial CF results
in the flat slope between lines B and C (see Fig. 2 (a)). The
increasing contribution of the 1st-3rd scattering 
as well as the emerging contribution from
the 4th scattering are responsible for the further slope increase after
the point B. Summing up our estimations for  $\delta p_{\bot}$ and
$\delta p_{\parallel}$
due to scattering, initial and asymptotic CF, the exact momentum change is
reproduced, see Fig. 4. Neglecting scattering,  initial or asymptotic CF, respectively,
the contribution of each effect is quantified. The multiple scattering
is crucial for the transverse momentum change. It determines the $\delta p_{\bot}(\varphi_i)$ behavior which, in turn, determines the ionization phases corresponding to the LES
peaks, and 
in this way 
the shape of LES. The initial CF plays a less important
role for the transverse momentum change, but a very significant
one for the longitudinal momentum change. The latter shifts the LES peak, determining
the position of LES. These two points are consistent with the
conclusions from Fig. 1(b). The asymptotic CF is only important at
near zero energies and has little impact on LES.

\begin{figure}[t]
\includegraphics[width=9 cm]{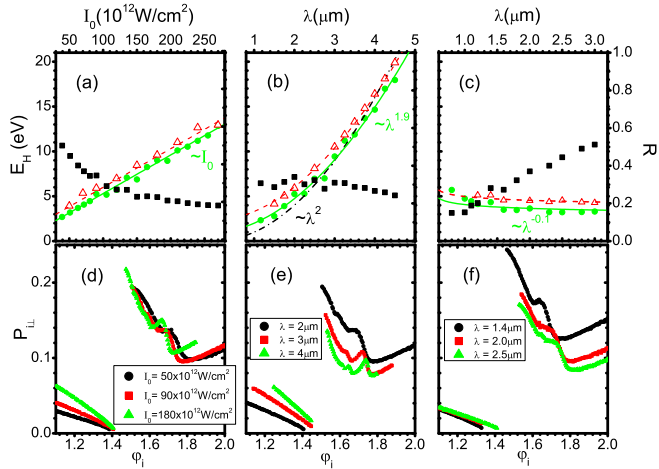}
\caption{(color online). The dependence of $E_H$ (circles) and $R$
(squares) on (a) laser intensity $I_0$ at $\lambda = 2\mu m$, (b)
wavelength $\lambda$ at $I_0 = 9.0 \times 10 ^{13}$ W/cm$ ^2$ and (c)
wavelength $\lambda$ at $\gamma = 0.534$. The solid line is a fit
to all of the LES data. The triangles and dashed lines are our estimations.
In (d), (e) and (f) the electron distribution in phase space for
different laser intensities and wavelengths are displayed. } \label{Fig_5}
\end{figure}

Finally, we investigate the  dependence of LES on the laser intensity
and wavelength, see Fig. 5. Two parameters characterize the LES, the first
is $E_H$, the high-energy limit of LES
\cite{DiMauroNP09}, as labeled in Fig. 1(b). The second
is the contrast ratio ($R$), the ratio between the integral yield of
photoelectrons in the energy interval
$(0,E_H)$ and the
total ionization yield for electrons emitted along the laser
polarization direction. $R$ characterizes the relative height and $E_H$
the width of the LES, respectively.
$E_H$ increases approximately linearly with
intensity (see Figs. 5 (a)).
Increasing the wavelength,
$E_H$ slightly deviates from the $\lambda^{2}$-law
 (i.e. the dependence on only $\gamma$) 
which is mainly due to the increasing initial CF contribution to the
$\delta p_{\parallel}$.
The $R$ dependence on the laser
parameters is more interesting. The $R$ decreases
monotonically with increasing intensity. With increasing
wavelength, $R$ first keeps constant and then decreases slowly.
The $E_H$ and $R$ behavior can be explained qualitatively by inspecting the phase space, see Figs. 5 (d), (e).
In Figs. 5(d) and (e), the curves in phase space rotate counter-clockwise resulting in the reduction
of electrons within the LES.
The reason is the competition between the initial CF and scattering. 
When fixing the Keldysh parameter (see Fig. 5(c),(f)),
the curves in phase space shift parallel to larger phase which yields to the $R$ increase with rising wavelength. At smaller wavelength, $R$, i.e. the LES
visibility, becomes smaller. This partly explains why the LES has
not been experimentally observed with near infrared lasers. Another
reason is the relative suppression of multiple scattering at
near infrared wavelength due to quantum effects \cite{chaotic}.

In conclusion, the experimental results have been successfully
reproduced via the semiclassical calculations and the origin of
LES is clarified.  At mid-infrared wavelengths the multiple
rescattering of ionized electron plays a significant nonperturbative role
which is the main factor creating the LES. The peaks in LES arise due to multiple scattering
contributions to the transverse momentum.

We gratefully acknowledge C. I. Blaga and L. F. DiMauro for
sharing the experimental data and C. H. Keitel  for
valuable discussions.

\newcommand{\noopsort}[1]{} \newcommand{\printfirst}[2]{#1}
\newcommand{\singleletter}[1]{#1} \newcommand{\switchargs}[2]{#2#1}


\begin{thebibliography}{99}



\bibitem{Becker_review} W. Becker \textit{et al.}, Adv. Atom. Mol. Opt. Phys. \textbf{48}, 36 (2000).

\bibitem{CorkumPRL93} P. B. Corkum, Phys. Rev. Lett. {\bf 71}, 1994 (1993).

\bibitem{Lein} M. Lein,  J. Phys. B \textbf{40}, R135 (2007).

\bibitem{SFA} L. V. Keldysh, Sov. Phys. JETP {\bf 20}, 1945 (1964); F. H. M. Faisal, J. Phys. B {\bf 6}, L89 (1973); H. R. Reiss, Phys. Rev. A {\bf 22}, 1786 (1980).

\bibitem{Bauer}  S. V. Popruzhenko \textit{et al.}, Phys. Rev. A \textbf{77}, 053409 (2008); I. A. Burenkov \textit{et al.}, Laser Phys. Lett. \textbf{7}, 409 (2010).

\bibitem{UllrichPRL03} R. Moshammer \emph{et al.}, Phys. Rev. Lett. {\bf 91}, 113002 (2003); A. Rudenko \emph{et al.}, J. Phys. B {\bf 37}, L407 (2004).

\bibitem{interference_structures} A successful quantum mechanical explanation is given by 
F. H. M. Faisal and G. Schlegel, J. Phys. B.  {\bf 38}, L223 (2005).

\bibitem{EichmannNature09} T. Nubbemeyer \emph{et al.},
Phys. Rev. Lett. \textbf{101}, 233001 (2008).

\bibitem{Gallagher} E. S. Shuman \emph{et al.}, Phys. Rev. Lett. \textbf{101}, 263001 (2008).


\bibitem{MidInfrared09} C. Erny \emph{et al.}, Appl. Phys. B {\bf 96}, 257 (2009).

\bibitem{DiMauroNP09} C. I. Blaga \textit{et al.},
Nature Phys. {\bf 5}, 335 (2009); F. Catoire \emph{et al.}, Laser
Phys. {\bf 19}, 1574 (2009).

\bibitem{XuPRL09} W. Quan \emph{et al.}, Phys. Rev. Lett. {\bf 103}, 093001 (2009).

\bibitem{KFR2} G. Duchateau \emph{et al.},
Phys. Rev. A {\bf 63}, 053411 (2001).

\bibitem{PaulusPRL94} The backwards scattering would
contribute to the plateau of the PES, see G. G.
Paulus \emph{et al.}, Phys. Rev. Lett. {\bf 72}, 2851 (1994).

\bibitem{FaisalNP09} F. H. M. Faisal, Nature Phys. {\bf 5}, 319 (2009).


\bibitem{N_s} The number of recollision can be defined as $N_s\sim  \omega l_m/2\pi v_d$, where $l_m$ is the  the maximal  drift distance and $v_d$ the drift velocity. During the drift in the laser polarization direction $v_d\sim \sqrt{2\varepsilon}$ with the  electron energy $\varepsilon$ and $l_m\sim \alpha$ because the electron passes the origin only when the drift distance is less than $\alpha$.  Due to the electron drift in the transversal direction, $v_d\sim p_{\bot}$. Accordingly,  $N_s\sim (\alpha \omega /2\pi)\times \min\{1/p_{\bot}, 1/\sqrt{2\varepsilon}\}=\min\{\alpha/x_0,(1/2\pi)\sqrt{2U_p/\varepsilon}\}$.

\bibitem{BurgPRA04} K. I. Dimitriou \emph{et al.}, Phys. Rev. A {\bf 70}, 061401(R) (2004).

\bibitem{RussiaLP09} N. I. Shvetsov-Shilovski \emph{et al.},
Laser Phys. {\bf 19}, 1550 (2009).


\bibitem{ADK}  A. M. Perelomov, V. S. Popov, and V. M. Teren'ev,
Sov. Phys. JETP {\bf 23}, 924 (1966); M. V. Ammosov, N. B. Delone, and V. P. Krainov, \emph{ibid}.
{\bf 64}, 1191 (1986); N. B. Delone and V. P. Krainov, J. Opt. Soc. Am. B {\bf 8}, 1207 (1991). 

\bibitem{Landau77} L. D. Landau and E. M. Lifshitz, \emph{Quantum Mechanics} (Pergamon, Oxford, 1977) p. 293.

\bibitem{Brabec} T. Brabec et al., Phys. Rev. A \textbf{54}, R2551 (1996).

\bibitem{chaotic} The contribution to LES of electrons  with chaotic dynamics increases with
 decreasing laser wavelength. However, at rather small laser
wavelength, when the impact parameter at first scattering becomes
comparable with the electron de Broglie wavelength: $p_{\bot}/\omega \lesssim
1/\sqrt{2\varepsilon}$,  the classical description fails. Then, the
diffraction of the electron wavepacket makes multiple scattering
inefficient (see also J. Tate \emph{et al.}, Phys. Rev. Lett.
{\bf 98}, 013901 (2007)) which suppresses the LES. For a LES energy
$\varepsilon \approx 3$ eV, $p_{\bot}=0.1$, the above condition
is $\lambda \lesssim 1$ $\mu$m.

\bibitem{IvanovPRA01} G. L. Yudin and M. Y. Ivanov, Phys. Rev. A {\bf 63}, 033404 (2001);
D. Comtois \emph{et al.}, J. Phys. B {\bf 38}, 1923 (2005); C. Huang \emph{et al.}, Opt. Express {\bf 18}, 14293 (2010).


\end{thebibliography}
\end{document}